\documentclass[aps,twocolumn,letterpaper,showpacs,10pt]{revtex4-1}
\usepackage[utf8]{inputenc}
\usepackage{graphicx,amsmath}
\usepackage{amssymb}
\usepackage{amsthm}

\begin{document}
\title{Synthetic Helical Liquids with Ultracold Atoms in Optical Lattices}
\author{J.\ C.\ Budich$^{1,2}$, C.\ Laflamme$^{1,2}$, F.\ Tschirsich$^{3}$, S.\ Montangero$^{3}$, P.\ Zoller$^{1,2}$}
\affiliation{$^1$Institute for Theoretical Physics, University of Innsbruck, 6020 Innsbruck, Austria}
\affiliation{$^2$Institute for Quantum Optics and Quantum Information, Austrian Academy of Sciences, 6020 Innsbruck, Austria}
\affiliation{$^3$Institute for Complex Quantum Systems \& Center for Integrated Quantum Science and Technology, University of Ulm, Albert-Einstein-Allee 11, D-89069 Ulm, Germany}

\date{\today}
\begin{abstract}
We discuss a platform for the synthetic realization of key physical properties of helical Tomonaga Luttinger liquids (HTLLs) with ultracold fermionic atoms in one-dimensional optical lattices. The HTLL is a strongly correlated metallic state where spin polarization and propagation direction of the itinerant particles are locked to each other.
We propose an unconventional one-dimensional Fermi-Hubbard model which, at quarter filling, resembles the HTLL in the long wavelength limit, as we demonstrate with a combination of analytical (bosonization) and numerical (density matrix renormalization group) methods. An experimentally feasible scheme is provided for the realization of this model with ultracold fermionic atoms in optical lattices. Finally, we discuss how the robustness of the HTLL against back-scattering and imperfections, well known from its realization at the edge of two-dimensional topological insulators, is reflected in the synthetic one-dimensional scenario proposed here. 
\end{abstract}
\maketitle

\section{Introduction}
The rich interplay between orbital degrees of freedom, spin, and many-body correlations gives rise to fascinating phenomena in quantum physics. New possibilities for their realization and observation are provided by the flexibility and control of quantum systems based on ultracold atoms in optical lattices \cite{ZwergerReview,LewensteinReview,ColdFermionReview}. Along these lines, synthetic magnetic fields  \cite{JakschZollerNJP,GerbierDalibard,DalibardReview} or even effects of non-Abelian gauge fields such as spin orbit coupling (SOC) \cite{spielmanReview} have been observed in systems consisting of neutral atoms \cite{Aidelsburger2011,Aidelsburger2013,Spielman2015,Marie2015}. As we show below, this recent  experimental progress even makes available the natural ingredients for the realization of quantum many-body systems which, while being inspired by intriguing concepts from condensed matter physics such as helical Tomonaga Luttinger liquids (HTLLs) \cite{XuMoore,Wu2006}, are not known to have a direct counterpart in real materials.

In the context of strongly correlated one-dimensional (1D) Fermi gases, a new physical twist has been provided by the discovery of the HTLL \cite{XuMoore,Wu2006}. In contrast to a conventional Tomonaga Luttinger liquid \cite{Tomonaga1950,Luttinger1963,MattisLieb,Haldane1981}, the HTLL is characterized by  a peculiar locking of spin and direction of motion of the itinerant particles, namely that fermions with opposite spin move in opposite direction. The HTLL has been originally predicted as a metallic edge theory of two-dimensional topological insulators \cite{KaneMele2005a,KaneMele2005b,BHZ2006, koenig2007} exhibiting the quantum spin Hall (QSH) effect. Due to their robust spin sensitive transport properties, HTLLs are promising candidates for numerous spintronics applications. In our present work, we propose and study an exotic 1D Fermi-Hubbard model that shows crucial aspects of HTLL physics at long distances, and is amenable to a systematic study and comparison to QSH edge states. The microscopic realization of this model in 1D optical lattices draws intuition from engineered gauge fields \cite{JakschZollerNJP,GerbierDalibard,DalibardReview,spielmanReview}. Reminiscent of SOC, our model contains a strongly spin-dependent hopping which, however, breaks time reversal symmetry, thus going conceptually beyond natural SOC. This unconventional feature is crucial for the basic physics characterizing the HTLL, namely that particles at opposite Fermi points have exactly opposite spin.\\

\begin{figure}[t!]
\includegraphics[width=0.98\columnwidth]{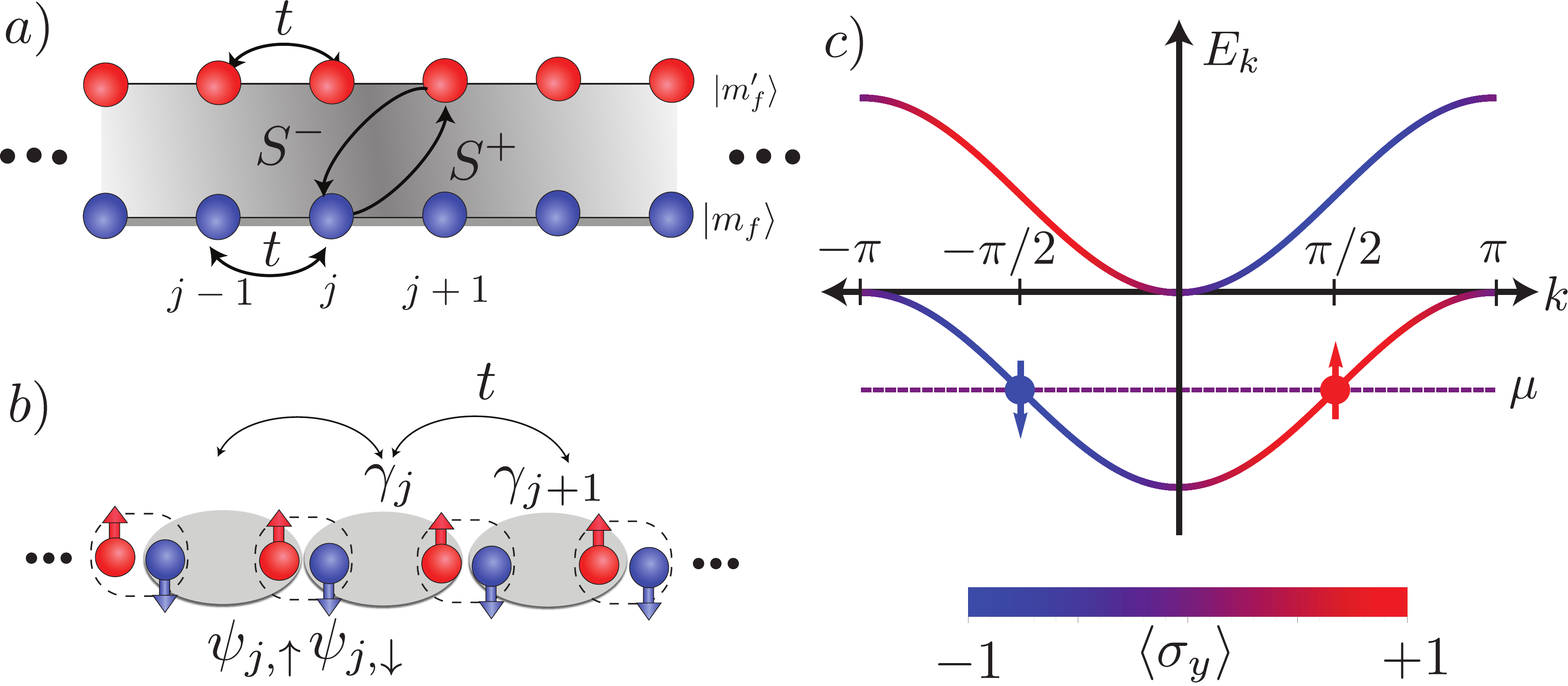}
\caption{ a) Schematic of the two-component model Hamiltonian~\eqref{eqn:tb} for $b=\alpha_R=1$, with hopping $t$ between sites of the same component, and operations $S^{+(-)}$ associated with a change in component and a hopping from left (right) to right (left), respectively.  b) Schematic of model projected onto the lower band, resulting in an effective single component model. Grey ovals denote operators  $\gamma_j$ of the effective nearest neighbor hopping model in the lower band (c.f. Eq.~\eqref{eqn:tblow}). Physical lattice sites are denoted by dashed lines. c) Band structure of the Hamiltonian (\ref{eqn:tb}) for $t=b=\alpha_R=1$ and chemical potential $\mu$ at quarter filling. The coloring of the plots visualizes the momentum-dependent spin polarization of the fermions, where red denotes spin up and blue spin down.}
\label{fig:bands}
\end{figure}

The systems we investigate may be considered synthetic in a double sense. First, they are inspired by edge states of 2D QSH systems but are purely 1D systems that do not rely on the presence of an insulating 2D bulk. Second, they are realized in synthetic material systems based on optical lattices in contrast to their semiconductor based counterparts. 
The motivation of our study is hence also twofold. Regarding the first point, it is well known that a system which has no other degrees of freedom than a single helical channel cannot exist in a 1D lattice system \cite{XLReview2010}. We would therefore like to address both qualitatively and quantitatively the question as to which aspects of HTLL physics can be seen in a purely 1D lattice system, and how much of the robustness against imperfections known from QSH systems survives in this synthetic scenario. Second, the experimental accessibility of subtle spin-dependent phenomena in HTLLs is limited in condensed matter systems by the presence of unavoidable imperfections, partly related to the surrounding bulk degrees of freedom. By contrast, in synthetic systems based on cold atoms in optical lattices, the tunability of individual parameters and accessibility of observables with single site resolution may open up possibilities to measure, e.g., the spin correlation functions of the single channel HTLL.\\

Below, we present an experimentally feasible scheme for the realization of our exotic 1D Fermi Hubbard model with $^{173}$Yb  atoms. The spin $1/2$ degree of freedom of the particles in our model is encoded in two hyperfine levels with different magnetic quantum numbers $m_f$. To engineer the hopping of the atoms, we use so called Raman assisted tunnelling techniques \cite{JakschZollerNJP,GerbierDalibard,DalibardReview}. %In this framework, the motion of atoms in the lattice is only energetically resonant in combination with a Raman process. 
This allows us to imprint laser phases, and by means of dipole selection rules, also non-Abelian spin operations on the hopping processes.
Beyond this experimental proposal, we extensively analyze our model theoretically. In the presence of on-site Hubbard interaction we assess, with a combination of analytical and numerical methods, in what sense its low-energy physics mimics a HTLL by computing long distance spin-spin correlation functions. Quite remarkably, the interacting model has a parameter line where these observable quantities can be exactly calculated analytically. Furthermore, we address the natural question to what extent the robustness against back-scattering - that can be understood as a topological protection in the natural realization of the HTLL as an edge state of a 2D topological insulator \cite{KaneMele2005a,Wu2006,BHZ2006,XLReview2010} - carries over to our synthetic one-dimensional scenario.\\

{\emph{Outline -- }} The remainder of this article is organized as follows. In Section \ref{sec:modelbuilding}, we discuss crucial features of the HTLL in general and introduce the microscopic lattice model that is at the heart of our present analysis. Thereafter, in Section \ref{sec:implementation}, we demonstrate how our model can be experimentally realized using state of the art techniques
to control ultracold fermionic atoms in optical lattices. Section \ref{sec:sol} is dedicated to the investigation of the low-energy physics of our model using a combination of analytical and numerical quantum many-body methods. Finally, in Section \ref{sec:conclusion}, we present some concluding remarks and put our main findings into a broader context. 

\section{Model building}      
\label{sec:modelbuilding}
This section is concerned with the modelling of synthetic HTLLs. In Section \ref{sec:hallmarks}, we build a case for simulating HTLL physics in 1D lattice systems by summarizing the key observable features of the HTLL and comparing the situation of its realization as an edge state of a 2D topological insulator to synthetic one-dimensional scenarios. Subsequently, in Section \ref{sec:model} we introduce the 1D lattice model which forms the basis of our present study.

\subsection{Hallmarks of the HTLL}
\label{sec:hallmarks}
The ideal HTLL has several intriguing properties which distinguish it decisively from both a spinless and a spinful single channel Tomonaga Luttinger liquid (TLL). Most prominently, there is no phase space to write down a spin-independent impurity term as discussed for the ordinary TLL in Refs. \cite{KaneFisherPRL,KaneFisherPRB}. This is because of the absence of states with equal spin for particles moving in opposite directions. For edge states of the time reversal symmetry (TRS) preserving QSH state, this robustness can be pushed even further and precludes elastic single particle back-scattering due to any TRS preserving single particle term in the Hamiltonian \cite{BHZ2006,XLReview2010}.     
This makes the transport characteristics of the ideal HTLL quite unique and various subtle dissipative effects that can cause inelastic back-scattering even in the presence of TRS have been investigated \cite{Wu2006,Stroem2010b,Budich12,Schmidt2012,FrancoisRashba,JukkaPuddles}. In a QSH system with a fixed spin quantization axis, the stability of the holographic HTLL at finite interaction strength has been investigated from first principles \cite{MartinHTLLQMC}. 

Another distinctive feature of the HTLL are its characteristic, very anisotropic spin-spin correlations. For spin operators $S^{\perp}$ perpendicular to the spin quantization axis of the eigenstates at the Fermi momenta $\pm k_F$, they read for large distances $r$ as
\begin{align}
\langle S^{\perp}(r)S^{\perp}(0)\rangle\sim\cos(2 k_F r)\frac{1}{r^{2K}},
\label{eqn:ssperb}
\end{align}
i.e., they exhibit Friedel oscillations and follow a power law that depends on the correlation strength via the Luttinger parameter $K$. In contrast, for spin operators $S^\shortparallel$ parallel to this quantization axis, they decay as
\begin{align}
\langle S^\shortparallel(r)S^\shortparallel(0)\rangle \sim \frac{1}{r^2},
\label{eqn:sspar}
\end{align} 
i.e., with an interaction independent power-law and without Friedel oscillations. Note that these correlation functions are a direct consequence of, and probe for, the helical nature of the Fermi surface: First, Eq. (\ref{eqn:sspar}) reflects that $S^\shortparallel$ is a good quantum number around the individual Fermi points. This is because if $S^\shortparallel$ were to mix left and right moving particles at the Fermi surface, these correlations would exhibit a non-universal power law decay. Second, Eq. (\ref{eqn:ssperb}) implies that spin-flips of $S^\shortparallel$, as described by the $S^{\perp}$-operators, do couple the two Fermi-points, hence excluding that both Fermi points have the same polarization direction of $S^\shortparallel$.\\

In non-holographic, i.e., purely 1D realizations of the HTLL (see, e.g., Refs.  \cite{Egger3DTI,vonOppenMaj,LossHelicalCNT,KloeffelHelical,HenrikNanowire} for various semiconductor based approximate realizations), there is a priori no topological protection against elastic single particle back-scattering by TRS and the extent to which the correlation functions concur with Eqs. (\ref{eqn:ssperb}-\ref{eqn:sspar}) has not been checked from first principles yet. In this work, we propose a minimal, experimentally feasible microscopic 1D lattice model where the helical nature of the Fermi points is an exact feature at quarter filling. 
Regarding the characteristic spin-spin correlations, we are able to demonstrate the presence of a correlated HTLL state in the sense of Eqs. (\ref{eqn:ssperb}-\ref{eqn:sspar}) in a finite parameter range by a combined numerical (in the framework of density matrix renormalization group (DMRG)) \cite{WhiteDMRG, SchollwoeckReview} and analytical (combination of mapping to an exactly solvable model and bosonization) approach. Furthermore, turning to the mentioned robustness against back-scattering, we find that a spin-independent fluctuation in the lattice potential in our model gives rise to a modified effective impurity term (see Section \ref{sec:robustness} below for a more detailed discussion). This may be interpreted as a certain robustness of our synthetic HTLL which is hence also expected to have quite characteristic transport properties as compared to the conventional TLL the further exploration of which is an interesting future direction. We argue how the situation of an ideal HTLL can be more closely mimicked by slightly complicating the underlying band structure. Even though our model globally breaks TRS, there is an emergent TRS at or around the Fermi surface which gives rise to  a certain robustness of transport properties.\\

\subsection{1D model for the synthetic HTLL}
\label{sec:model}

We consider a lattice model with a single spin-$\frac{1}{2}$~fermionic degree of freedom per site and choose the  lattice constant as our unit of length. The free tight-binding model of the underlying band structure is given by
\begin{align}
H_0=\frac{1}{2}\sum_j \psi^\dag_j \left[b \sigma_x + i \alpha_R \sigma_y -t\sigma_0\right]\psi_{j+1} +\text{h.c.}, 
\label{eqn:tb}
\end{align}
where $\psi_j=(\psi_{j,\uparrow},\psi_{j,\downarrow})^T$~are spinors of fermionic field operators, $t$~is the ordinary spin-independent hopping strength, $\alpha_R$~denotes the Rashba velocity, and $b$~tunes a TRS breaking hopping term which may be seen as an exotic Zeeman term. On Fourier transform we obtain the Bloch Hamiltonian
\begin{align}
&h_0(k)=d^\mu(k) \sigma_\mu,\quad\mu = 0,x,y,z\nonumber\\
&d^\mu(k)=(-t\cos(k),b \cos(k),\alpha_R \sin(k),0).
\label{eqn:BlochHam}
\end{align}
The band structure and the Bloch functions explicitly read as
\begin{align}
&E_{\pm}(k)=d^0\pm \lvert \vec d\rvert,~\vec d= (d^x,d^y,d^z);\nonumber\\
&\lvert u_{\pm}(k)\rangle=\frac{P_{\pm}(k)\lvert \uparrow\rangle}{\lvert P_{\pm}(k)\lvert \uparrow\rangle \rvert}, \quad \sigma_z \lvert \uparrow\rangle = \lvert \uparrow\rangle\nonumber\\
&P_{\pm}(k)=\frac{1}{2}\left(1\pm \hat d(k)\cdot\vec \sigma\right),~\hat d=\frac{\vec d}{\lvert \vec d\rvert}.
\label{eqn:freesol}
\end{align}
Due to the structure of $\vec d$, we immediately see that at $\pm k_F=\pm \frac{\pi}{2}$, i.e. at the Fermi points at quarter filling, the Bloch states of the lower band are characterized by $\sigma_y \lvert u_-(\pm k_F)\rangle=\pm \lvert u_-(\pm k_F)\rangle$, i.e., they are exact $\sigma_y$~eigenstates with opposite eigenvalue at the opposite Fermi points. This reflects the helical nature of the Fermi surface for a half filled lower band (quarter filling of the lattice).
In Fig. \ref{fig:bands}c), we show the band structure of the free model for $t=b=\alpha_R=1$~visualizing the lattice-momentum-dependent spin polarization of the Bloch states and indicating the helical Fermi points.\\

We emphasize the role of the term $b\cos(k) \sigma_x$ appearing in the Bloch Hamiltonian (\ref{eqn:BlochHam}). It differs crucially from an ordinary Zeeman term as induced by a magnetic field in a nano-wire (see, e.g., Ref. \cite{vonOppenMaj}) because it is momentum-independent. A Zeeman term $B\sigma_x$ opens a gap between the two spin species, even at $k=0$ where the Rashba term $\alpha_R \sin(k)\sigma_y$ vanishes. However, it also gives an equal $\sigma_x$ spin polarization to the Bloch states at opposite Fermi points thus competing with the desired helical nature of the Fermi surface. In contrast, the term $b\cos(k)\sigma_x$ appearing in Eq.~(\ref{eqn:BlochHam}) is maximal at $k=0$ but vanishes exactly at $k=\pm \frac{\pi}{2}$, i.e., at the Fermi points at quarter filling. The Rashba term being an odd function of momentum then gives rise to exactly helical Fermi points.\\

We note that $b \cos(k) \sigma_x$ could be replaced by $b \cos^n(k) \sigma_x$ in Eq.~(\ref{eqn:BlochHam}) at the expense of introducing $n$-nearest neighbor hopping in Eq.~\eqref{eqn:tb}. That way, not only the Fermi points but also an  expansion to order $\delta k^{n-1}$ around them would be exactly helical. For $n=2$, for example, the full linearized theory around the Fermi surface would give an exact HTLL up to band curvature terms. In the following, we focus on the case $n=1$, where the model is to some extent amenable to analytical study even in the presence of interaction.\\

Due to the absence of a stable Fermi liquid in 1D, considering interaction effects is of key importance to make experimentally relevant predictions on helical liquids. Here, we model the interaction between the fermionic atoms with an ordinary on site Hubbard interaction term,
\begin{align}
\label{eq:Int_Ham}
H_I = U \sum_j n_{j,\uparrow} n_{j,\downarrow},\quad n_{j,\sigma}=\psi_{j,\sigma}^\dag \psi_{j,\sigma},~\sigma=\uparrow, \downarrow
\end{align}
with interaction strength $U$. In Section \ref{sec:sol}, we show that the model described by 
\begin{align}
H=H_0+H_I
\label{eqn:ham}
\end{align}
indeed exhibits HTLL physics in a finite parameter range.

\section{Implementation with cold atoms in optical lattices}
\label{sec:implementation}
Intuition for the practical realization of our model (\ref{eqn:tb}) is drawn from both theoretical proposals \cite{JakschZollerNJP,GerbierDalibard} and recent experiments \cite{Aidelsburger2011,Aidelsburger2013,Spielman2015,Marie2015} on synthetic classical gauge fields in optical lattices, where Raman assisted tunnelling techniques are used to engineer the phase of the hopping amplitude of the atoms. The experimental scheme proposed here is related to the experimental setup in Ref. \cite{Marie2015}, where a magnetic flux in a so called synthetic dimension formed by internal states of $^{173}$Yb atoms has been realized. However, as we detail below, our present proposal goes crucially beyond this scheme since it contains a spin-flip hopping process, which would correspond to a combined hopping in the synthetic dimension (i.e. the spin flip), and hopping in the real dimension (see diagonal hopping terms in Fig. \ref{fig:bands}\,a).\\

We now detail how the tight-binding model (\ref{eqn:tb}) can be realized using Raman assisted tunnelling techniques in a system of ultracold fermionic atoms in a 1D optical lattice. Motivated by recent experimental progress \cite{Marie2015}, we focus on the fermionic alkaline earth atoms $^{173}$Yb. The ground state in this scenario is a $^1S_0$ state, with $F=I=\frac{5}{2}$ for  $^{173}$Yb, and we choose to form the spin $\sigma$ in Eq.~\eqref{eqn:tb} with $m_f=-5/2$ and $m_f'=-1/2$.
The occupation of the remaining states in the manifold $m_f=-5/2,\ldots, +5/2$ is inhibited by a combination of dipole selection rules and energy conservation arguments, as we will explain in more detail below. This effective projection onto the subspace forming $\sigma$ has already been experimentally achieved in Ref. \cite{Marie2015}. We note that the most relevant parameter regime for our proposal is defined by $b=\alpha_R$ in Eq. (\ref{eqn:tb}), where our model is also amenable to analytical study (see Section \ref{sec:sol}). 
However, our implementation will also allow for the case $b\ne \alpha_R$.  
We start by lifting the degeneracy of the six $m_f$-states by a Zeeman splitting $\Delta_z$ induced via a magnetic field in the $z$-direction.
Then, to implement the Hamiltonian Eq.~\eqref{eqn:tb} for $b=\alpha_R$, we need to engineer the following two nearest neighbor hopping terms (see Fig. \ref{fig:bands}a).
 \begin{enumerate}
 \item A spin-dependent hopping of strength $b=\alpha_R$, where hopping from left to right is accompanied by the spin operation $S_+=\frac{1}{2}(\sigma_x + i \sigma_y)$ while the hermitian conjugate $S_-$ acts when hopping from right to left.
 \item A spin-independent nearest neighbor hopping of strength $t$.
 \end{enumerate} 
 
We start by addressing the first term. This term is related to what has been demonstrated in \cite{Marie2015}, however again we emphasize the difference here is that the spin flip term (hopping in the synthetic dimension) is associated with nearest neighbour hopping (in the spatial dimension). 
%This term goes beyond Ref. \cite{Marie2015}, where a scheme consisting of spin-independent hopping and on-site terms changing the internal state of the atoms has been experimentally realized. The latter process is called a hopping in a synthetic dimension, physically realized as the mentioned $m_f$-states of the $^{173}$Yb ground state in Ref. \cite{Marie2015}. In Fig. \ref{fig:bands} a, such processes can be viewed as vertical hopping processes, i.e., on-site spin flips. We emphasize that the spin-flip hopping of strength $b=\alpha_R$ in our model, in contrast, corresponds to diagonal hopping terms, where the particle moves both in the real and in the synthetic dimension (see Fig. \ref{fig:bands}\,a). 
This additional ingredient is physically crucial here to achieve the exact helical polarization of the Fermi points in our model and is hence at the heart of our current proposal. To implement this specific nearest neighbor hopping term, the natural spin-independent hopping $J$ stemming from the overlap of the Wannier functions is prohibited by tilting the optical lattice with a slope $\Delta \gg J$. Several ways to achieve this have been reported \cite{JakschZollerNJP}. Here, we need to tilt the two spin species with an equal slope, which speaks against using a magnetic field gradient. Instead, ways of achieving this are to employ a spatial gradient in the laser intensity that may be realized by a laser focus, to "shake" the lattice by means of a time-dependent frequency difference of the laser beams, or to simply tilt the lattice by means of the gravitational potential. This can generate a slope in the optical lattice potential which is equal for our two spin species.

 \begin{figure}[t!]
\begin{center}
\includegraphics[width=0.8\columnwidth]{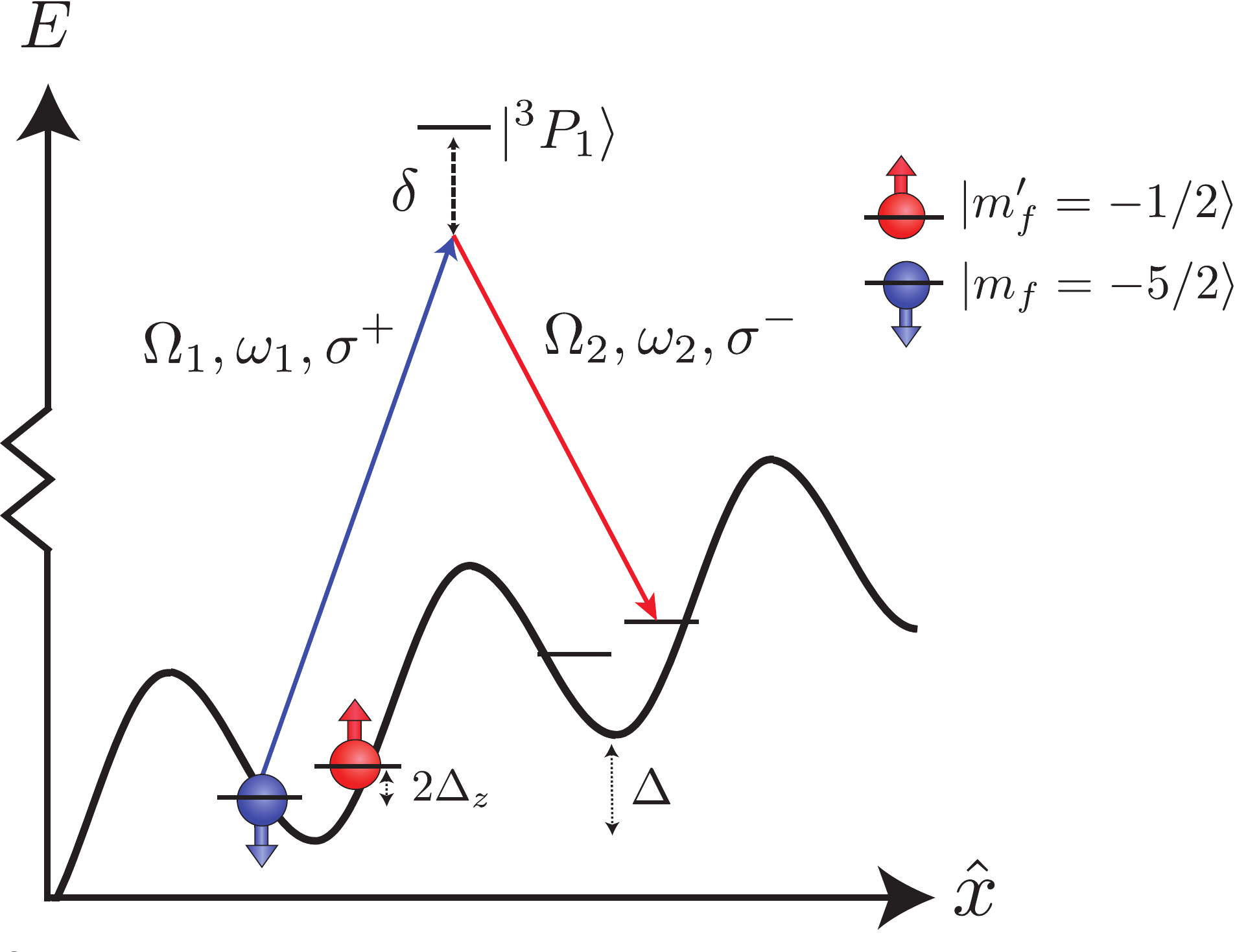}
\caption{Proposed setup for the implementation of the spin-dependent hopping in Eq.~(\ref{eqn:tb}) at the point $\alpha_R=b$ using $^{173}$Yb. The two spin states are given by the nuclear zeeman states of the $^1$S$_0$ ground state, $|m_f=-\frac{5}{2}\rangle$,$|m_f'=-\frac{1}{2}\rangle$, subject to a Zeeman splitting of $2\Delta_z$. Raman assisted tunneling with lasers with Rabi frequencies $\Omega_{1/2}$, frequencies $\omega_{1/2}$ and polarizations $\sigma^{+/-}$, respectively, couple nearest neighbor sites via a $|^3P_1\rangle$ virtual excited state, detuned by an amount $\delta$. The lattice is tilted with equal tilting $\Delta$ of the two spin states.}
\label{fig:level_scheme}
\end{center}
\end{figure}

The hopping is then restored in the tilted lattice with the help of two far detuned Raman lasers \cite{Aidelsburger2013}. The locking between the spatial hopping direction and the spin flip operators $S_\pm$, respectively, is achieved by dipole selection rules. More concretely, due to $m_{f'}-m_f=2 \hbar$, two units of angular momentum along the $z$-direction need to be transferred in every hopping process. To this end, we use a pair of Raman lasers with frequencies $\omega_1, \omega_2$, polarisations and wave vectors $\sigma_1=\sigma_+$, ${\bf{k_1}}=\frac{\omega_1}{c}{\bf{e_z}}$ and $\sigma_2=\sigma_-,~ {\bf{k_2}}=\frac{\omega_2}{c}{\bf{e_z}}$, respectively. The Raman lasers are detuned by $\delta$ from a manifold of excited states, here given by the $^3 P_1$ state of $^{173}$Yb.
For $^{173}$Yb, the hyperfine splitting between the excited states with different $F'=(\frac{7}{2},\frac{5}{2})$ is about $4.7$ GHz \cite{YbSpectroscopy}, so that the detuning $\delta$ of the Raman process may also be on the order of $1$ GHz \cite{Marie2015}, i.e., far detuned compared to both the energy scale of the optical lattice and the line-width of the corresponding optical transition. If the frequency difference concurs with the tilting and the Zeeman splitting as  $\omega_1-\omega_2=\Delta + 2 \Delta_z$, spin-flip nearest neighbor hopping assisted by a two photon Raman process conserves energy. 
As well, hopping processes corresponding to a climb (descend) in the tilted lattice are associated with distinct photon processes, namely $\omega_1$ is absorbed (emitted) and $\omega_2$ is emitted (absorbed), respectively. Due to the distinct polarizations $\sigma_1=\sigma_+,\sigma_2=\sigma_-$ of the photons, a climb (descend) in the lattice corresponds to a transfer in the angular momentum along the $z$-direction of $2\hbar$ ($-2 \hbar$), precisely matching the difference in the magnetic quantum numbers between our two spin species.
This establishes a one-to-one correspondence between the direction of the Raman assisted hopping and the application of the spin flip operators $S_\pm$ (see Fig.~\ref{fig:level_scheme}) and prevents the dynamical occupation of three of the other four $m_f$-states $m_f=-3/2,1/2,5/2$ that are not part of our spin $\sigma$. More concretely, $m_f=-3/2,1/2,5/2$ are energetically off-resonant by at least $\Delta_z$ and, in addition, cannot be reached from our spin states by transferring an even number of units of angular momentum along the $z$-direction. In order to prevent occupation of the fourth state, $m_f=3/2$, its energy is shifted away from resonance by a light shift \cite{Marie2015}.

In order to not occupy higher Bloch bands of the optical lattice, the hopping strength $\alpha_R$ resulting from our Raman assisted spin-flip tunnelling  scheme should be smaller than band gap $\Delta_b$ between the lowest and the second lowest Bloch band. Lastly, we note that unwanted onsite spin-flips, which would correspond to an ordinary, i.e., momentum-independent Zeeman term in (\ref{eqn:tb}) are off-resonant by the lattice tilting $\Delta$ and are hence strongly suppressed if $\alpha_R<\Delta$.

We point out that the model engineered so far, i.e., Eq.~(\ref{eqn:tb}) with $t=0$ at half filling realizes a flat band 1D topological insulator, similar to the model introduced by Su, Schrieffer, and Heeger \cite{SSH}. However, in order to obtain a finite Fermi velocity at the helical Fermi points at quarter filling, a finite $t$ is necessary to make the flat bands dispersive. This brings us to the second term above, the spin-independent hopping $t$. This can be realized by another two-photon Raman assisted tunnneling process which is not spin-selective. However, for this process, a spin-dependent phase may be implemented by varying the direction of the Raman lasers in order to give their wave vector a component parallel to the 1D optical lattice ($x$-direction). This allows us to tune away from $\alpha_R=b$, thus enabling the implementation of Eq.~(\ref{eqn:tb}) in all generality. 

Finally, we note that the helical nature of the Fermi points occurs at quarter filling of the lattice, i.e., at a particle density of one atom per two lattice sites on average. This commensurate filling could be achieved experimentally by, for example, temporarily switching on a superlattice with the double period and realizing a Mott phase with one particle per site in this super lattice.\\

\section{Analytical and numerical analysis}
\label{sec:sol}
Our Hamiltonian (\ref{eqn:ham}) represents a 1D Fermi-Hubbard model with a somewhat exotic underlying band structure (at $b=\alpha_R=0$~it would be the ordinary Hubbard model). The long wavelength physics of this model will be revealed with a combination of analytical and numerical methods in the following. 

\subsection{Exact solution and lower band projection}
As a first approach, we present an elegant way to map the low-energy physics around the Fermi points of this model to that of an exactly solvable model. We consider the lattice at quarter filling and put $b=\alpha_R, t=1$. We project the model to its lower band which gives a good approximation of the low-energy physics around the Fermi surface, if the energy scales of interest are much smaller than the energetic separation $2b-1$ of the Fermi surface from the upper band (see Fig. \ref{fig:bands} right panel). The effective free Hamiltonian then reads as
\begin{align}
\tilde H_0= \sum_k E_-(k)l_k^\dag l_k
\end{align}
where $l_k^\dag$~are the creation operators of the Bloch states in the lower band, i.e., $l_k^\dag \lvert 0\rangle = \lvert u_-(k)\rangle$. Now, we also project the interaction term $H_I$~to the lowest band. To this end it is helpful to have a Wannier basis of localized states spanning the lower band. The $k$-independence of $\lvert \vec d\rvert$~at $b=\alpha_R$~significantly helps here (see Eq.~(\ref{eqn:freesol})). Following Ref. \cite{Eddy1DFTP} and considering that the ordinary hopping term $t$~does not influence the Bloch states due to its spin-independence, we obtain
\begin{align}
\label{eqn:tblow}
&\tilde H_0=-\frac{1}{2}\sum_j\left[\gamma_j^\dag\gamma_{j+1}+\text{h.c.}\right],\\
&\gamma_j=\frac{1}{\sqrt{2}}\left(\psi_{j,\downarrow}-\psi_{j+1,\uparrow}\right)\nonumber,
\end{align}
where we have shifted the Energy by $b$ units to get rid of the constant term $\lvert \vec d(k)\rvert=b$~in the spectrum $E_-(k)=-\cos(k)-\lvert \vec d(k)\rvert$~of the lower band. The effective momentum-dependent model (\ref{eqn:tblow}) for the lower band is visualized in Fig. \ref{fig:bands}b). The projection $\tilde H_I$~of the interaction Hamiltonian $H_I$~to the lower band assumes in terms of the localized Wannier operators $\gamma_j$ the simple form
\begin{align}
\tilde H_I = \frac{U}{4}\sum_j \tilde n_j \tilde n_{j+1},\quad \tilde n_j = \gamma_j^\dag\gamma_j.
\label{eqn:ulow}
\end{align}
Putting together Eq.\,(\ref{eqn:ham}-\ref{eqn:ulow}), we have
\begin{align}
\tilde H=& \mathcal P_- H \mathcal P_- =\tilde H_0+\tilde H_I= \nonumber \\
&-\frac{1}{2}\sum_j\left[\gamma_j^\dag\gamma_{j+1}+\text{h.c.}\right]+\frac{U}{4}\sum_j \tilde n_j \tilde n_{j+1},
\label{eqn:effmod}
\end{align}
where $\mathcal P_-$~denotes the projection onto the lower band. Eq.~(\ref{eqn:effmod}) represents an effective spinless fermion model in terms of the operators $\gamma_j$ with only nearest neighbor interaction which is commonly referred to as the $t-V$ model. However, via the spin structure of the $\gamma_j$ operators (see Eq. (\ref{eqn:tblow})), a spin dependence is encoded in this effective spinless model. Here, $t=1$ and $U/4$ plays the role of $V$. This model can be solved exactly using the Bethe ansatz and is, at half filling, known to describe a Tomonaga Luttinger liquid (TLL) for interactions $U<4$. At $U>4$~the system develops a $2k_F$~charge density wave order and a gap is opened. Note that the quarter filling of the full lattice we started with now corresponds to half filling of the effective one-band model.\\

\subsection{Bosonization and characteristic spin correlations}
Our goal is to verify that the long distance physics of our model indeed exhibits the characteristic behaviour of the HTLL. The most striking signature of the HTLL is the  asymptotic decay of the spin-spin correlation functions which depends drastically on the spin direction.
The spin-spin correlations perpendicular to the polarization axis of the eigenstates (here the $y$-axis) decay with a non-universal power law that depends on the Luttinger parameter $K$ and exhibit Friedel oscillations (see Eq.~(\ref{eqn:ssperb})). In contrast, the spin-spin correlations parallel to this axis decay with the second power in distance, independent of the interaction strength, and do not show Friedel oscillations (Eq.~(\ref{eqn:sspar})).
From Eq.~(\ref{eqn:freesol}) we know that the $\lvert u_-(\pm k_F)\rangle$~at $\pm k_F=\pm \frac{\pi}{2}$~describe perfectly helical modes indicating that the long wavelength physics of our model resembles the HTLL. However, in the presence of interaction, long range correlation functions of the projected model in Eq.~(\ref{eqn:ulow}) are hard to access from its exact Bethe ansatz solution.
Still, we know from the exact solution that the low-energy theory of the effective spinless model (\ref{eqn:effmod}), is a spinless TLL the correlation functions of which are analytically computable. We hence project the microscopic lattice spin operators $S^i_j,~i=x,y,z$ at site $j$ to the lower band and bosonize them in a long wavelength continuum model that is linearized around the Fermi energy. To this end, we first decompose the Fermi operators $\gamma_j$ of the lower band into a right-moving and a left moving part
\begin{align}
\gamma_j = \psi_L(x)+\psi_R(x),~x=j
\label{eqn:decomp}
\end{align}
and treat $x$ as continuous parameter in the subsequent analysis. Following the notation of Ref. \cite{GiamarchiBook}, the bosonized form of the operators reads in the thermodynamic limit as
\begin{align}
\psi_p(x)= \frac{U_p}{\sqrt{2\pi \alpha}}\text{e}^{i p k_F x}\text{e}^{-i(p \phi(x)-\theta(x))}, \quad p=R/L=\pm, 
\label{eqn:bosonization}
\end{align}
where $U_p$ are the mutually anti-commuting Klein factors, $\alpha$ is a short distance cutoff, and $\phi, \theta$ are the bosonic phase field and its dual, respectively.
The main merit of the bosonized representation is that the interacting model (\ref{eqn:effmod}), linearized around the Fermi energy, is quadratic in the bosonic fields which allows for the analytical calculation of the long range correlation functions (see, e.g., Ref. \cite{GiamarchiBook}). To employ this for the calculation of the desired spin-spin correlation functions, we project the lattice spin operators $S_j^i=\psi_{j,\alpha}^\dag\frac{\sigma^{\alpha\beta}_i}{2}\psi_{j,\beta},~i=x,y,z$ to the lower band. Looking at Eq.~(\ref{eqn:tblow}), it is clear that this amounts to the mapping
\begin{align}
\psi_{j,\downarrow}\rightarrow \frac{1}{\sqrt{2}}\gamma_j,~\psi_{j,\uparrow}\rightarrow -\frac{1}{\sqrt{2}}\gamma_{j-1}.
\label{eqn:fermilbp}
\end{align}
Along with Eqs. (\ref{eqn:decomp}), (\ref{eqn:bosonization}), the spin operators can now in principle be brought into bosonized form. However, from Eq.~(\ref{eqn:fermilbp}), it is clear that the off-diagonal spin operators $S^i_j,~ i=x,y$, contain lowest band operators $\gamma_j, \gamma_{j-1}$ at neighboring sites. Explicitly, 
\begin{align*}
S^x_j \rightarrow -\frac{1}{4}(\gamma_j^\dag \gamma_{j-1}+\gamma_{j-1}^\dag \gamma_j),~S^y_j\rightarrow \frac{i}{4}(\gamma_{j-1}^\dag \gamma_j-\gamma_j^\dag \gamma_{j-1}).
\end{align*}

In the linearized continuum model, by virtue of Eq.~(\ref{eqn:decomp}), operators like $\psi_p^\dag(x-1)\psi_{p'}(x),~p,p'=L,R$ hence appear which are separated by the short distance of $1<\alpha$. 
In order to evaluate the relevant correlation functions using the bosonic long wavelength theory such terms first need to be simplified to operators evaluated at a single position by performing an operator product expansion (OPE). Terms with $p=p'$ above give leading contributions proportional to $\partial_x \phi$ or $\partial_x \theta$. Pair correlations evaluated at positions $x$ and $y$ of such terms give rise to the universal power-law $\lvert x-y\rvert^{-2}$ appearing in Eq.~(\ref{eqn:sspar}). In contrast, the leading OPE of terms with $p\ne p'$, e.g., $\psi_R^\dag(x-1)\psi_{L}(x)$ gives rise to operators that oscillate with $2 k_F x$ and that depend on the bosonic fields as $\text{e}^{i2\phi(x)}$. Pair correlations of such operators give rise to the non-universal power-law $\lvert x-y\rvert^{-2K}$ appearing in Eq.~(\ref{eqn:ssperb}) and their oscillatory behavior causes the concomitant Friedel oscillations. The explicit evaluation of the long distance spin-spin correlators $\langle S^i(x)S^i(y)\rangle$ with $\lvert x-y\rvert \gg 1,\alpha$ is now tedious but straight forward. For the parallel $\langle S^y(x)S^y(y) \rangle$ correlations, relative signs in the matrix structure of the spin operators lead to a cancellation of all oscillatory terms, hence resulting indeed in the universal power law and the absence of Friedel oscillations displayed in Eq.~(\ref{eqn:sspar}). For the perpendicular spin-spin correlations, in contrast, the leading correlations at repulsive interactions ($K<1$) are produced by the oscillatory terms which cause the $2k_F$ Friedel oscillations and give rise to the non-universal power-law appearing in Eq.~(\ref{eqn:ssperb}). The bosonization analysis hence confirms that the spin-spin correlation functions of our model at quarter filling and $b=\alpha \gg t$ decay with the characteristic behavior of the HTLL. At $K=1/2$ two particle umklapp scattering terms become relevant and the system enters a charge density wave phase as predicted by the Bethe ansatz solution.

\subsection{Numerical analysis}
Complementary to the previous analytical approach, we treat the full microscopic lattice model (\ref{eqn:ham}) in the framework of finite system DMRG \cite{SchollwoeckReview}. This allows for a direct measurement of the correlation functions along the directions parallel and perpendicular to the polarization at the Fermi surface. More specifically, we measure various spin-spin correlation functions at positions $r_1$ and $r_2$ and fit their decay of  to the power-law $r^{-2 \xi}$ with $r=\lvert r_1 -r_2\rvert$. To minimize finite size effects we keep $r_1$ and $r_2$ at fixed fractions of the system length $L$, namely $r_1=3 L/4,~r_2=L/4$ and vary the distance $r=L/2$ by variation of the system size $L$  \cite{MartinHTLLQMC}. Besides confirming our analytical analysis from first principles, our numerical study also allows us to access a broader parameter range where the model is not amenable to analytical treatment.\\
\begin{figure}[t]
\includegraphics[width=0.95\columnwidth]{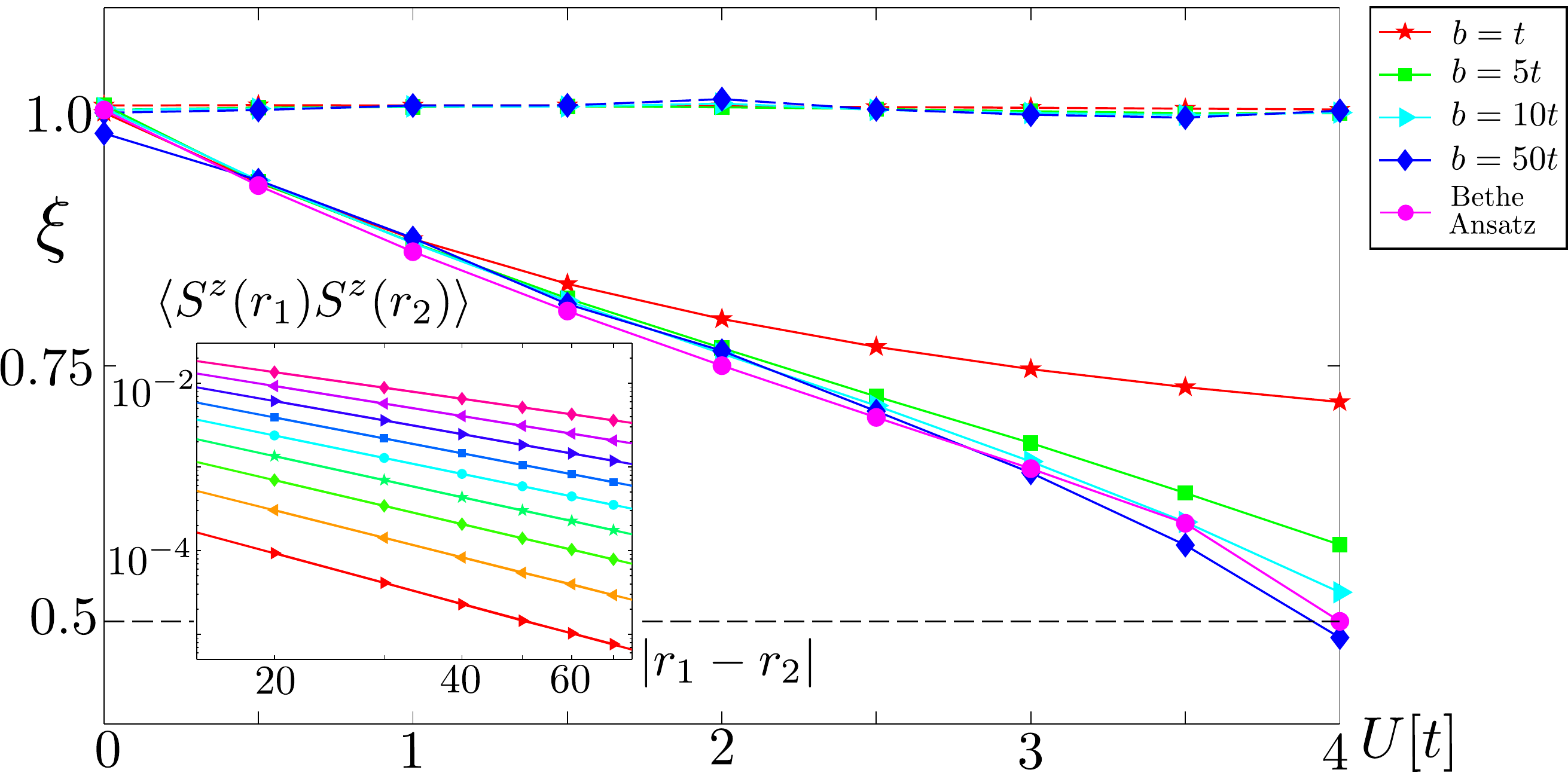}
\caption{The decay exponent $\xi$ extracted from a fit to $|r_1-r_2|^{-2\xi}$ as a function of the Hubbard interaction $U$, as extracted from DMRG measurements of the correlation function $\langle S^z(r_1)S^z(r_2)\rangle$ (solid) and $\langle S^y(r_1)S^y(r_2)\rangle$ (dashed), respectively. Here, $r_1=3L/4$ and $r_2=L/4$, where the system size $L$ is varied between $40$ and $140$ sites. Different colors correspond to separation of the bands increasing from $b=\alpha_R=t$ to $b=\alpha_R=50t$. Additionally, the solution of the Bethe Ansatz is shown for comparison. The value of $\xi$=0.5 indicating the phase transition between the Luttinger liquid and charge-density wave phase is shown with a dashed black line as a guide for the eye. The spread around $\xi=1$ at $U=0$ gives an indication of the magnitude of the numerical error. {\it Inset:}  Correlation function $\langle S^z(r_1)S^z(r_2)\rangle$ as a function of $|r_1-r_2|$, and the best fit from which the exponent $\xi$ can be extracted. Here all data is at $\alpha_R=b=10t$, and the colors indicating $U=0$ increasing to $U=4t$. }
\label{fig:KvsU}
\end{figure} 
We first concentrate again on the case $b=\alpha_R$.
For the parallel spin direction, we find that the correlations are basically independent of interaction, confirming the behavior given in Eq.~\eqref{eqn:sspar}. By contrast, the correlation functions in the perpendicular direction depend heavily on the interaction strength, in agreement with Eq.~(\ref{eqn:ssperb}). The dependence of the Luttinger parameter $K$ on the interaction strength $U$ can be extracted from finite size scaling of the perpendicular $\langle S^z(r_1)S^z(r_2)\rangle$ correlations, where $K=\xi$. Numerical errors from our DMRG calculations and finite size effects limit the precision of the fits for $\xi$ to a precision of a few percent. The results are shown in Fig.~\ref{fig:KvsU}, for separation of bands increasing from $b=\alpha_R=t$ to $b=\alpha_R=50t$. When the band separation is much larger than the interaction strength, the projection to the lower band is a good approximation, and the Luttinger parameter follows the behavior of the Bethe Ansatz \cite{Karrasch2012} within the numerical errors of our finite size scaling. In this limit, the system crosses a phase transition into a charge-density wave phase at $U=4$; the critical point is at a Luttinger parameter of $K=0.5$ \cite{White2000}. However, when the interaction becomes comparable to the band separation, projecting out the upper band is no longer a good approximation as the interaction can readily mix population in both bands, and the behavior of the system deviates systematically from the prediction of the Bethe Ansatz.\\

%\begin{figure}[t]
%\includegraphics[width=0.95\columnwidth]{direct_correlations}
%\caption{The Luttinger parameter $K$ as a function of the Hubbard interaction $U$, as extracted from DMRG measurements of the correlation function $\langle S^z(r)S^z(0)\rangle$. Different colors correspond to separation of the bands increasing from $b=\alpha_R=t$ to $b=\alpha_R=100t$. Additionally, the solution of the Bethe Ansatz (valid until $U=4$) is shown for comparison. The value of K=0.5 is shown with a dashed black line as a guide for the eye. }
%\label{fig:KvsU}
%\end{figure}

\begin{figure}[t]
\includegraphics[width=0.95\columnwidth]{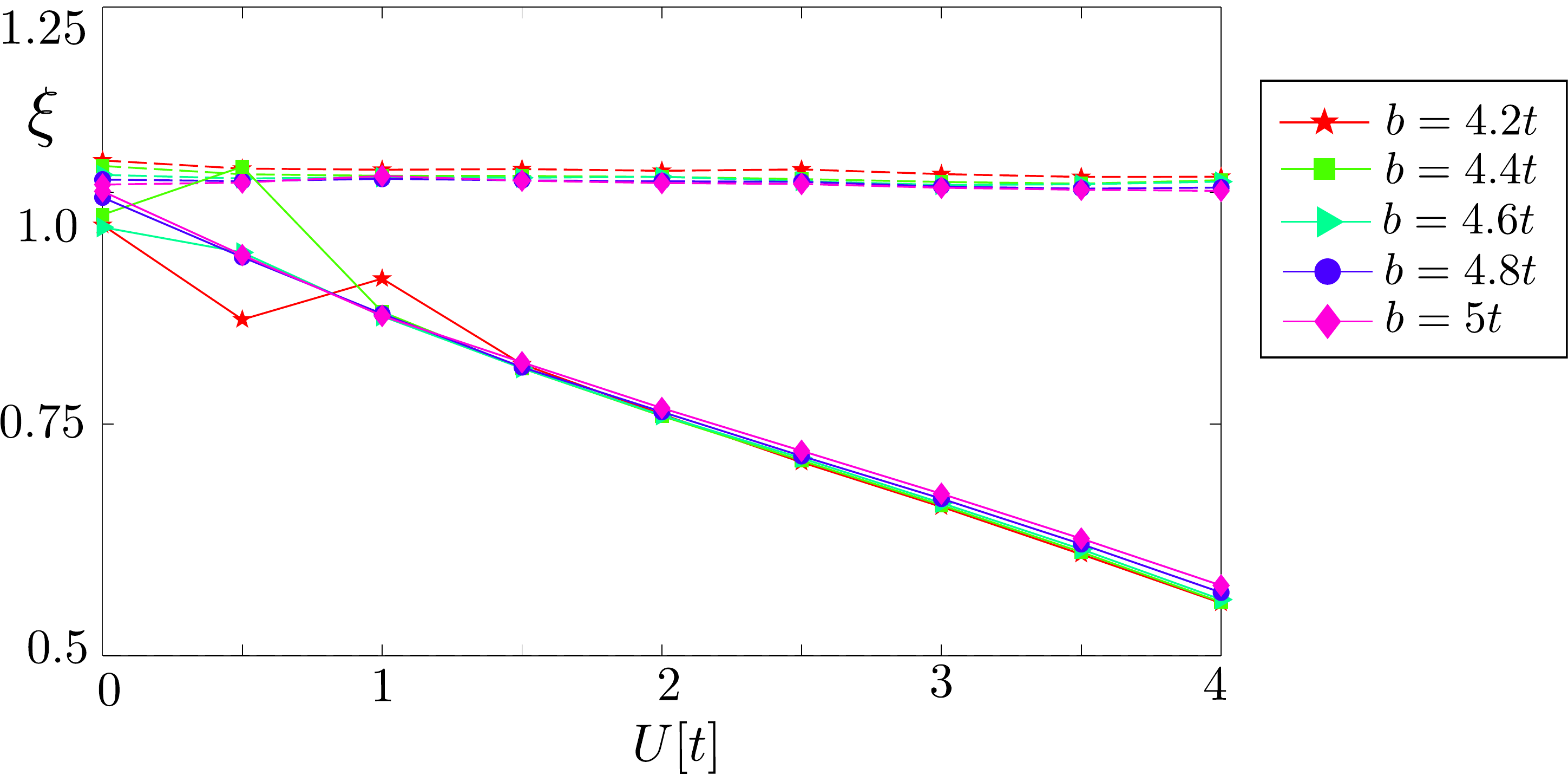}\\
\includegraphics[width=0.95\columnwidth]{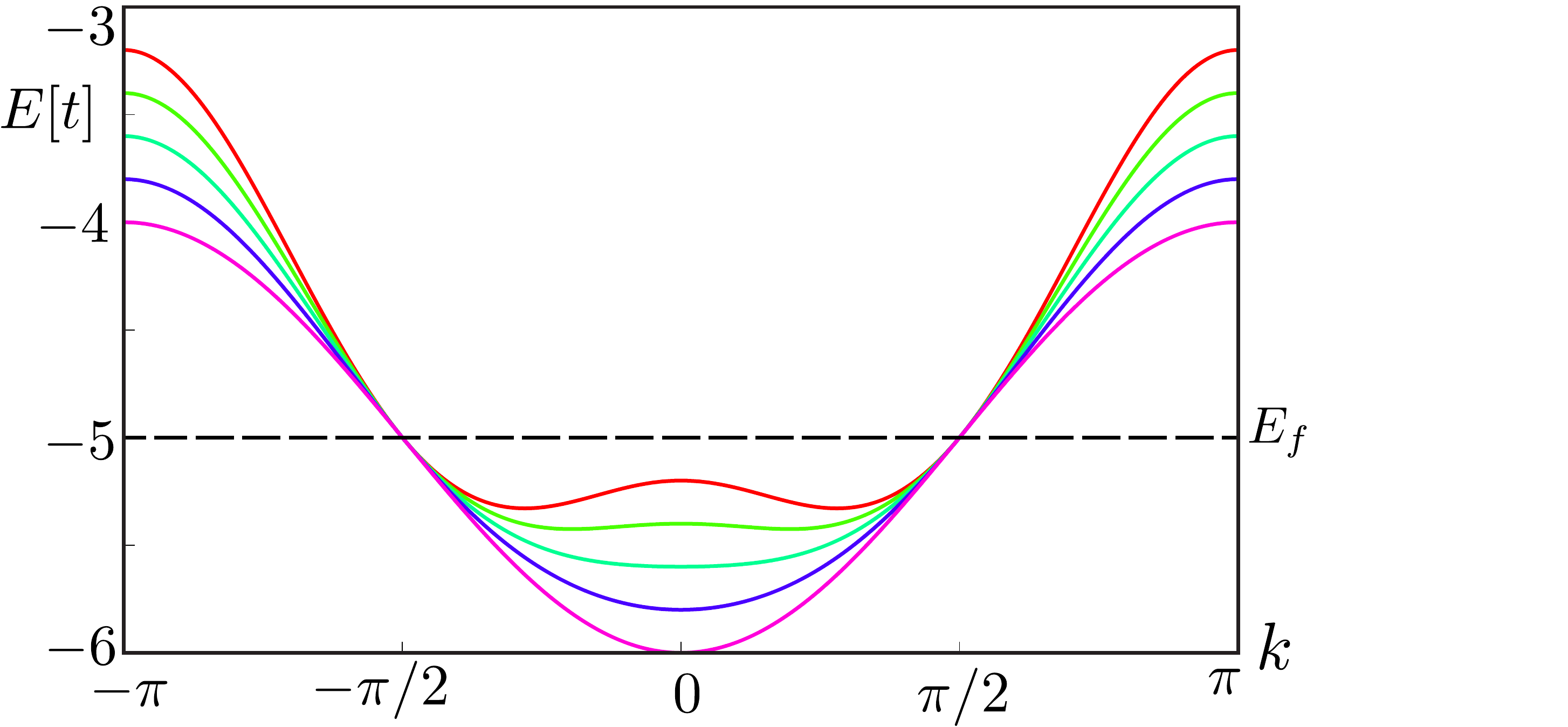}
\caption{{\it Upper Panel:}  Decay exponent $\xi$  from fit to $|r_1-r_2|^{-2\xi}$ as a function of the Hubbard interaction $U$, as extracted from DMRG measurements of the correlation function parallel  $\langle S^\shortparallel(r_1)S^\shortparallel(r_2)\rangle$ (dashed) and perpendicular $\langle S^\perp(r_1)S^\perp(r_2)\rangle$ (solid) to the spin quantization axis.
Here, $r_1=3L/4$ and $r_2=L/4$, where the system size $L$ is varied between $40$ and $140$ sites.
 Different colors correspond to $b=4.2t$ to $b=5t$, with $\alpha_R=5t$ fixed for every value of $b$. {\it Lower Panel:} Band structure of the lower band for the parameters of the upper panel.}
\label{fig:KvsU_unequal}
\end{figure}

We now consider the case $b \ne \alpha_R$, where the dispersion of the lower band is not just a simple $-t\cos(k)$ but is influenced by the spin-dependent hopping terms (see Eq.~(\ref{eqn:freesol})). As a consequence, the lowest band projection is no longer given by the exactly solvable model (\ref{eqn:effmod}). However, as long as $\lvert b-\alpha_R\rvert <t$, there are still only two exactly helical Fermi points at quarter filling. In the upper panel of Fig.~\ref{fig:KvsU_unequal}, we show the $U$-dependence in the decay exponent $\xi$ of the parallel and perpendicular spin-spin correlation functions $|r_1-r_2|^{-2\xi}$ for fixed $\alpha_R=5t$ and various values of $b$. The concomitant dispersion of the lower band for each value of $b$ is shown in the lower panel of Fig.~\ref{fig:KvsU_unequal}. The qualitative structure of the correlation functions is still similar to the case $b \ne \alpha_R$, in particular the striking anisotropy between the parallel and the perpendicular correlations. Discrepant points in the exponent of the perpendicular correlation function appear when the difference $\alpha_R - b$  becomes relevant when compared to the interaction strength. However, when the interaction is strong enough with respect to $\alpha_R-b$, the behavior is similar to that of the case $\alpha_R=b$. We hence conclude that the HTLL long distance behavior of our model persists also in an extended parameter regime where it is not exactly solvable.  

\subsection{Robustness against back-scattering}
\label{sec:robustness}
A hallmark of the HTLL is its robustness against single particle back-scattering. For the ideal HTLL consisting of nothing but a right moving branch with spin up and a left moving branch with spin down conjugated by TRS, elastic single particle back-scattering cannot be induced by any TRS preserving perturbation. In particular, a generic spin-independent impurity term causing back-scattering in an ordinary TLL cannot exist since the spin needs to be flipped in order to couple the opposite branches. In contrast, our present model (\ref{eqn:ham}) is a microscopic lattice model with a spin degree of freedom for which any local operator, irrespective of its spin dependence can be written down. The question that we address in this Section is to what extent phenomena reminiscent of this protection against back-scattering occur in our synthetic model, which, away from the Fermi surface, deviates from the ideal HTLL.\\

The simplest conceivable imperfection, a spin-independent impurity with strength $V_0$ at site $j=0$ can be microscopically modeled by the operator
\begin{align}
O_{\text{imp}}(j=0)=V_0 \psi^\dag_0\psi_0=V_0(\psi_{0,\uparrow}^\dag \psi_{0, \uparrow}+\psi_{0,\downarrow}^\dag \psi_{0, \downarrow}).
\label{eqn:imp}
\end{align}
The Fermi points at $\pm k_F=\pm \frac{\pi}{2}$ have exactly opposite spins and thus the matrix element for scattering from $k_F$ to $-k_F$ vanishes.
The admixture of the opposite spin polarization away from the Fermi surface scales linear with the deviation $q=k-k_F$.
To assess the effect of $O_{\text{imp}}$ on the low-energy theory around the Fermi surface, we first perform a projection to the lowest band.
With $\tilde n_j=\gamma_j^\dag \gamma_j$ and $\gamma_j=\frac{1}{\sqrt{2}}\left(\psi_{j,\downarrow}-\psi_{j+1,\uparrow}\right )$ (see Eq. (\ref{eqn:tblow})), we obtain
\begin{align}
\tilde O_{\text{imp}}(j)=\frac{V_j}{2} (\tilde n_j+\tilde n_{j-1}).
\end{align}
Using Eq. (\ref{eqn:decomp}), we translate $\tilde O_{\text{imp}}$ into the long wavelength continuum theory, which gives
\begin{align}
&H_{\text{imp}}=\sum_j \tilde O_{\text{imp}}(j)=\sum_j \frac{V_j}{2} (\tilde n_j+\tilde n_{j-1}) \approx \\
&\int \text{d}x \,V(x)\left [ \rho(x) + \left(\text{e}^{-i2k_F x}\partial_x \tilde \psi_R^\dag(x)\tilde \psi_L+\text{h.c.}\right)\right],\nonumber
\end{align}
where we defined $\rho(x)=\tilde \psi_L^\dag(x)\tilde \psi_L(x)+\tilde \psi_R^\dag(x)\tilde \psi_R(x)$ and $k_F=\frac{\pi}{2}$ has been used.
In bosonized form, the potential term $H_{\text{imp}}$ reads as
\begin{align*}
\int \text{d} \, x V(x) \left[ \frac{-\partial_x \phi}{\pi}+\frac{1}{2\pi \alpha}\left(\partial_x\phi\text{e}^{-i(2k_F x-2\phi)}+\text{h.c.}\right) \right],
\end{align*}
where the first term describes forward-scattering and the second term back-scattering. Compared to the conventional 
spinless TLL, where the back-scattering of a potential is described by a term  $\sim V(x)\cos(2 \phi(x))$  \cite{KaneFisherPRL, KaneFisherPRB}, there is an additional pre-factor of $\partial_x \phi(x)$ in our model. Simple power counting would hence indicate that back-scattering is less relevant in our model. However, to higher order in perturbation theory, the combination of forward-scattering and back-scattering may also generate the conventional term $\sim \cos(2 \phi(x))$ in our model, but with leading order $V^2$ in the potential strength. Thus, in the perturbative regime, where temperature and/or bias voltage are larger than the potential strength $V$, back-scattering is expected to be suppressed in our model compared to a conventional spinless TLL.  

We note that if we modify our model (\ref{eqn:tb}) by substituting $b \cos^2(k)$ for $b\cos(k)$ (see discussion in Section \ref{sec:model}), the impurity operator given in Eq. (\ref{eqn:imp}) becomes less relevant in the effective many-body theory and enters only via band curvature terms which go beyond the linearized TLL model.\\

We stress that the above discussion assumes that the Fermi wave vector is tuned exactly to $k_F=\frac{\pi}{2}$. In practice, the Fermi energy may slightly deviate from this value, say $k_F-\frac{\pi}{2}=\delta$. This will give rise to a conventional impurity term $\sim \delta V(x) \cos(2 \phi(x))$ on the order of this deviation. However, if the bias voltage $U_B$ or the temperature $T$ are larger than $\delta V$, this term will not pinch off the conductance. In conclusion, for $\delta V, V^2<U_B,T<V$, a scattering potential of strength $V$ would pinch off a conventional TLL while the synthetic HTLL is only perturbatively affected. In this sense, the experimental verification of the reduced back-scattering discussed here does not require fine tuning of the Fermi momentum.

\section{Concluding remarks}
\label{sec:conclusion}
We have demonstrated how HTLL physics naturally emerges in a synthetic framework based on ultracold fermionic atoms in optical lattices. In this context, static spin-spin correlation functions have been calculated as a hallmark of HTLL physics. These observable quantities are experimentally accessible in state of the art experiments on optical lattices and provide a probe of the helical nature of the Fermi surface.  Furthermore, the robustness of the ideal HTLL against back-scattering has been shown to have an interesting counterpart in our synthetic realization. In a broader context, the study of mesoscopic transport properties of neutral atoms in optical lattices is an emerging field of research (see, e.g., Ref. \cite{EsslingerScience2012}). Along these lines, our present work may open up a playground for the study of intriguing transport properties in the framework of unconventional quantum impurity problems. Once the single channel regime becomes accessible in transport experiments on ultracold atoms in optical lattices, HTLL physics in the proposed setup may also be dynamically probed by observing the correlation between the spin polarization and propagation direction of excitations.\\

\section*{Acknowledgments}
We acknowledge interesting discussions with Eddy Ardonne and Pietro Silvi. CL is partially supported by NSERC. We acknowledge financial support from the ERC synergy grant UQUAM  and from EU via SIQS as well as the DFG via the SFB/TRR21. This work was supported by the Austrian Ministry of Science BMWF as part of the UniInfrastrukturprogramm of the Focal Point Scientific Computing at the University of Innsbruck.\\

{\emph{Note added. --
When preparing this manuscript for submission, we became aware of the preprint Barbarino et al. arXiv:1504.00164, where evidence of helical liquids has been reported in a model with spin-independent hopping which maps to a Rashba spin orbit coupled wire.}}

\bibliographystyle{apsrev}

\end{document}